\documentclass[prb,aps,tightenlines,twocolumn,reprint,
showpacs,floatfix]{revtex4}
\usepackage{graphicx}
\usepackage{amssymb}
\usepackage{amsmath}
\usepackage{bm}
\usepackage{colordvi}
\usepackage{mathrsfs}
\usepackage{cases}

\begin{document}
\title{Anomalous D'yakonov-Perel' spin relaxation in semiconductor
 quantum wells under strong magnetic field in Voigt configuration}
\author{Y. Zhou}
\author{T. Yu}
\author{M. W. Wu}
\thanks{Author to whom correspondence should be addressed}
\email{mwwu@ustc.edu.cn.}
\affiliation{Hefei National Laboratory for Physical Sciences at
  Microscale and Department of Physics, University of Science and
  Technology of China, Hefei, Anhui, 230026, China}

\date{\today}
\begin{abstract}
We report an anomalous scaling of the D'yakonov-Perel' spin relaxation
with the momentum relaxation in semiconductor quantum
wells under a strong magnetic field in the Voigt configuration.
We focus on the case that the external magnetic field is perpendicular to the
spin-orbit-coupling--induced effective magnetic field and its magnitude is much
larger than the later one. 
It is found that the longitudinal spin relaxation time 
is proportional to the momentum relaxation time even in the 
strong scattering limit, indicating that the D'yakonov-Perel' spin
relaxation shows the Elliott-Yafet-like behavior.
Moreover, the transverse spin relaxation time is 
proportional (inversely proportional) to the momentum relaxation time in
  the strong (weak) scattering limit, both in the opposite trends against the 
well-established conventional D'yakonov-Perel' spin relaxation behaviors. 
We further demonstrate that all the above anomalous scaling relations come from
the unique form of the effective inhomogeneous broadening.
\end{abstract}



\maketitle

\section{Introduction}
In recent years, semiconductor spintronics has aroused enormous interest
due to the potential application of spin-based
devices.\cite{opt-or,spintronics,wu_review} 
Among intensive works in this field, the spin relaxation, 
which describes the decay of the out-of-equilibrium 
spin polarizations,  in n-type
semiconductor quantum wells (QWs) is an important area. 
The relevant spin relaxation mechanisms in this system are the 
Elliott-Yafet\cite{EY_original} (EY) and the D'yakonov-Perel'\cite{DP_original} 
(DP) mechanisms. 
In the EY mechanism, electron spins have a small chance to flip during each
scattering due to spin mixing. 
Thus the spin relaxation time (SRT) $\tau_s$ is proportional to the momentum
relaxation time $\tau_p$, i.e., $\tau_s\propto\tau_p$.
In the DP mechanism, electron spins decay due to their precession around the 
momentum-dependent effective magnetic field (which gives a dynamic analogue of
the inhomogeneous broadening\cite{wu_review,IB}) 
induced by the Dresselhaus\cite{Dresselhaus_55}
and/or Rashba\cite{Rashba_84} spin-orbit coupling (SOC) $\bm{\Omega}({\bf k})$
during the free flight between adjacent scattering events.
In the strong scattering limit, i.e., $\langle|\Omega({\bf k})|\rangle \tau_p \ll 1$
with $\langle...\rangle$ denoting the ensemble average, 
the DP spin relaxation satisfies the relation 
$\tau_{s}^{-1}\sim \langle \Omega^2({\bf  k}) \rangle \tau_p$,\cite{opt-or}
indicating that the SRT is inversely proportional to $\tau_p$. By contrast, in the weak
scattering limit, i.e., $\langle|\Omega({\bf k})|\rangle \tau_p> 1$, the DP SRT is
proportional to $\tau_p$.\cite{opt-or}
In most cases, the strong-scattering criterion is satisfied,\cite{wu_review}
and the distinct momentum-scattering-time dependence of the DP and EY SRTs is widely
used to distinguish which mechanism dominates the spin relaxation in the
experiments in semiconductors\cite{wu_review} and more recently in 
graphene.\cite{Wees_exp,Kawakami_exp,Avsar_exp,Kettemann_exp,Zhang_12_njp}

However, most of previous works only investigate the spin relaxation with zero
or weak magnetic fields. 
In this paper, we show the anomalous scaling of the DP spin relaxation
with the momentum relaxation under a strong magnetic field which is parallel
to the QW plane (the Voigt configuration), perpendicular to the spin-orbit
field and satisfies the condition
$\omega_B=g\mu_BB\gg \langle|\Omega({\bf k})|\rangle$.
A typical system satisfying the above conditions is a symmetric (110) QW with 
small well width.\cite{Ohno_110,Wu_110,Dohrmann_04,spin_noise_110,Sherman_110} 
The Hamiltonian  can be written as ($\hbar\equiv 1$ throughout this paper) 
\begin{align}
  H=\sum_{{\bf k}\sigma\sigma'}\left\{\varepsilon_{\bf k}\delta_{\sigma\sigma'}
    +\left[ g\mu_B{\bf B} +\bm{\Omega}({\bf k})\right] \cdot 
    \frac{\bm{\sigma}_{\sigma\sigma'}}{2}
    \right\} c^{\dagger}_{{\bf k}\sigma} c_{{\bf k}\sigma'}+H_{\rm I}.
\end{align}
Here $\varepsilon_{\bf k}=k^2/2m^\ast$ is the kinetic energy of electron with
momentum ${\bf k}=(k_x,k_y)$; $\bm{\sigma}$ are the Pauli matrices;
\begin{equation}
  \bm{\Omega}({\bf k})=\frac{\gamma_Dk_x}{2}(0, \; 0, \; k_x^2-2k_y^2- 
  \langle k_z^2\rangle_0 )
\end{equation}
is the effective magnetic field from the Dresselhaus\cite{Dresselhaus_55} SOC, 
with $\gamma_{\rm D}$ denoting the Dresselhaus SOC coefficient and
$\langle k_z^2 \rangle_0$ standing for the average of the operator
$-(\partial / \partial z)^2$ over the electronic state of the lowest
subband. The interaction Hamiltonian $H_{\rm I}$ is composed of
the electron-electron, electron-phonon and electron-impurity interactions.
Without loss of generality, we choose ${\bf B}$ along the $y$-axis.
In this situation, the SRTs along different directions can be expressed as 
\begin{align}
 \tau_{sz}^{-1}=\tau_{sx}^{-1}= \left\langle
    a\frac{\overline{\Omega_z^2({\bf k})}^2}{4\omega_B^2}\tau_p 
    + b\frac{{\Omega_z^2({\bf k})}}{\omega_B^2\tau_p} \right\rangle,& 
  \nonumber \\
  {\rm when}\ 
  \langle \frac{\overline{\Omega_z^2({\bf k})}}{2\omega_B} \rangle
  \ll\tau_p^{-1}\ll\omega_B; &
  \label{SRT_z_rough} \\
  \tau_{sy}^{-1} = \left\langle b' \frac{2{\Omega_z^2({\bf k})}}
    {\omega_B^2\tau_p} \right\rangle, \quad {\rm when}\ \tau_p^{-1}\ll\omega_B.&
  \label{SRT_y_rough} 
\end{align}
Here $a$, $b$ and $b'$ are the coefficients (around 1) depending on the specific  
momentum scattering mechanism; $\overline{A_{\bf k}} = A_{\bf k} - 
\frac{1}{2\pi}\int_0^{2\pi} d \phi_{\bf k}\,A_{\bf k}$.
We first address the transverse SRT perpendicular to the magnetic field, i.e., 
Eq.~(\ref{SRT_z_rough}). In the regime 
$\tau_p^{-1}\ll\sqrt{\frac{a}{4b}\langle\overline
{\Omega_z^2({\bf k})}^2\rangle/\langle{\Omega_z^2({\bf k})}\rangle}\sim
\langle|\Omega({\bf k})|\rangle$,
corresponding to the original (i.e., $B=0$) weak scattering limit,
the first term in Eq.~(\ref{SRT_z_rough}) is dominant and thus
$\tau_s\propto\tau_p^{-1}$. This indicates that the DP spin relaxation in the
original weak scattering limit exhibits the strong scattering behavior.
In the regime $\tau_p^{-1}\gg\sqrt{\frac{a}{4b}\langle\overline
{\Omega_z^2({\bf k})}^2\rangle/\langle{\Omega_z^2({\bf k})}\rangle}$,
corresponding to the original strong scattering limit, the second term in
Eq.~(\ref{SRT_z_rough}) is dominant and hence $\tau_s\propto\tau_p$, indicating
that the DP spin relaxation shows exactly the EY-like behavior.
Both behaviors are in the opposite trend against the conventional DP
ones. 
Thus we refer to these two regimes as the anomalous DP- and
EY-like regimes in the following.
For the longitudinal SRT parallel to the magnetic field, i.e.,
Eq.~(\ref{SRT_y_rough}), it is shown that $\tau_s\propto\tau_p$ even in the
original strong scattering limit, similar to the anomalous EY-like regime 
for the transverse SRT.

The paper is organized as follows: In Sec.~II, we discuss the 
effective inhomogeneous broadening and reveal the physics under the
anomalous DP behavior. In Sec.~III, we present the analytic formulae
and numerical results of the SRTs from the kinetic spin Bloch equation (KSBE)
approach. We conclude and discuss in Sec.~IV.

\begin{figure}[tbp]
  \begin{center}
    \includegraphics[width=8.cm]{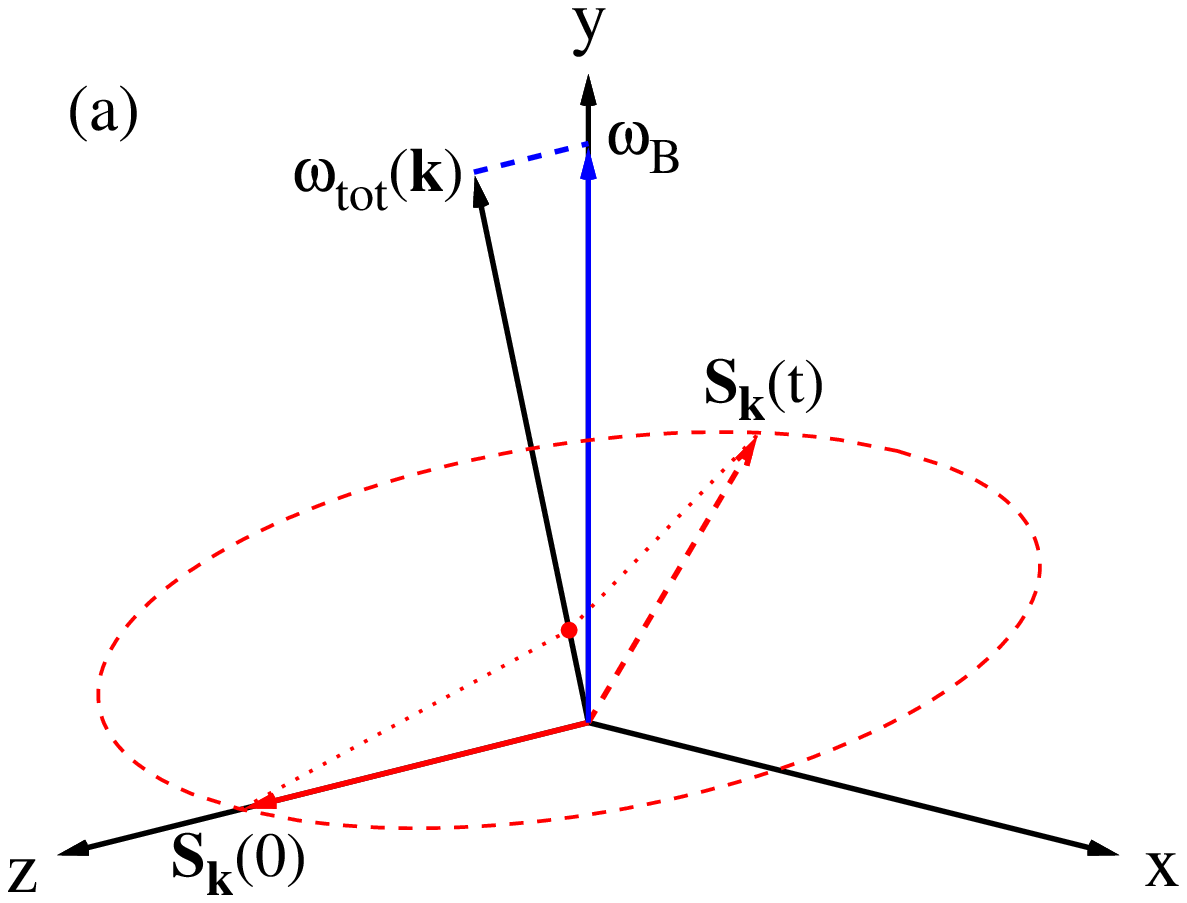}
    \includegraphics[width=8.cm]{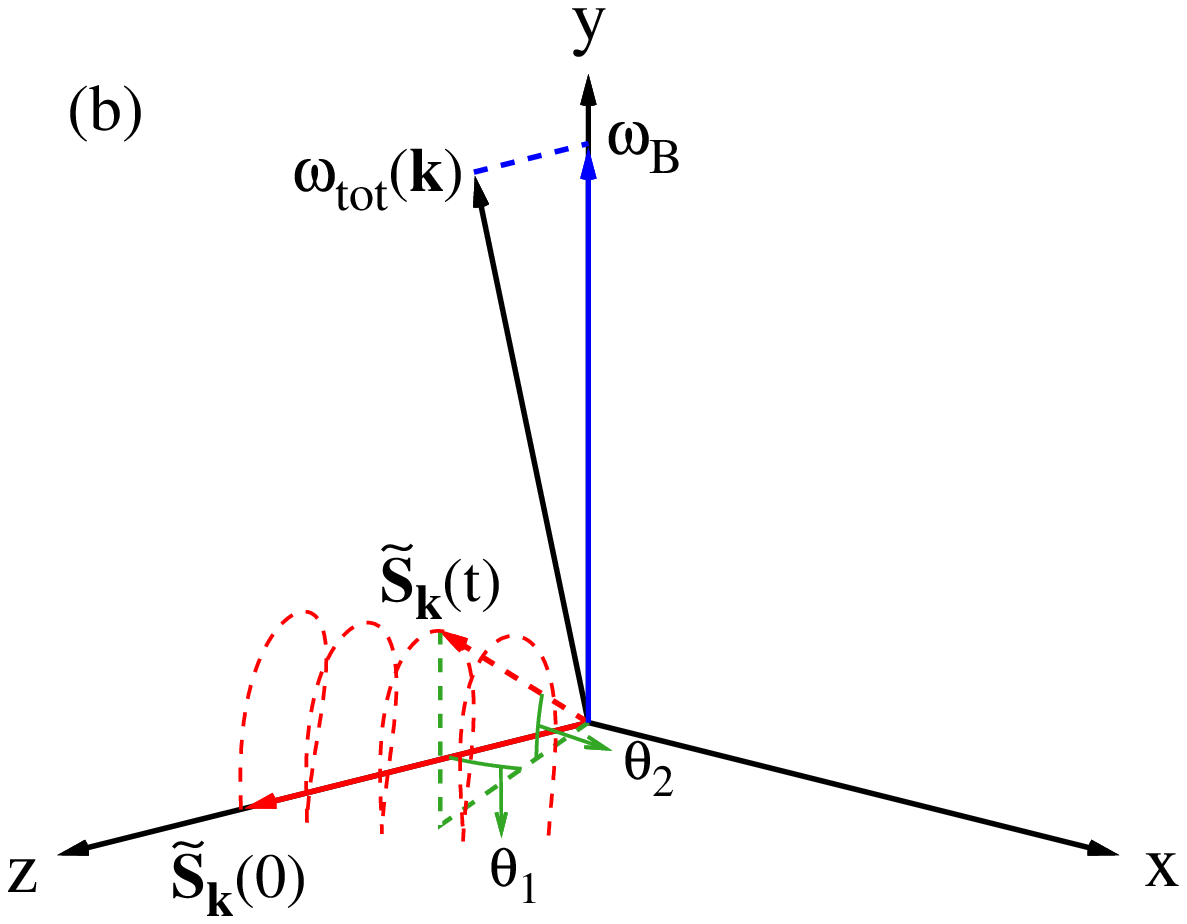}
  \end{center}
  \caption{(Color online) 
 The schematic illustration of the free spin precession 
    in the case with the initial spin polarization along the $z$-axis in the
    Schr\"odinger (a) and interaction (b) pictures. 
    The red dashed curves indicate the precession orbits.
    The red dotted lines in (a) connect the spin vectors 
    ${\bf S}_{\bf k}(0)$ and ${\bf S}_{\bf k}(t)$
    with the center of its precession orbit.
    Here we have exaggerated the angle between 
    $\bm{\omega}_{\rm tot}({\bf k})$ and
    the $y$-axis in order to make the relevant precession angles
 more pronounced. 
  }
  \label{fig_scheme} 
\end{figure}

\section{Effective inhomogeneous broadening}
To reveal the physics under the anomalous scaling of the DP SRT, we first
  discuss the effective inhomogeneous broadening by analysing
  the free spin precession between adjacent scattering events.
Without scattering, the spin vector ${\bf S}_{\bf k}(t)$ just precesses around
the total magnetic field $\bm{\omega}_{\rm tot}=\omega_B{\bf e}_y
+\Omega_z({\bf k}) {\bf e}_z$. Then one obtains 
\begin{equation}
{\bf S}_{\bf k}(t)=R_{\rm tot}(\omega_{\rm tot}({\bf k})t){\bf S}_{\bf k}(0),
\end{equation}
in which $R_{i}(\phi)$ represents the rotation operator
with angle $\phi$ around the direction ${\bf e}_{i}$. 
We take the case with the initial spin vector ${\bf S}_{\bf k}(0)$ along the
$z$-axis as a typical example to schematically show the precession orbit 
in Fig.~\ref{fig_scheme}.
It is seen that the main contribution of the precession angle is from the strong
external magnetic field, which is momentum independent and hence does not
contribute to the inhomogeneous broadening.
To show the effective inhomogeneous broadening more clearly,
we transform the spin vector into the interaction picture as
$\tilde{\bf S}_{{\bf k}}(t) = R_y(-\omega_B t) {\bf S}_{{\bf k}}(t)$, 
whose orbit is also plotted in Fig.~\ref{fig_scheme}.
Then the spin evolution operator in the interaction picture $U_{\bf k}(t,0)$, 
defined as $\tilde{\bf S}_{{\bf k}}(t)=U_{\bf k}(t,0)
\tilde{\bf S}_{{\bf k}}(0)$, can be obtained as
\begin{eqnarray}
  \nonumber
  U_{\bf k}(t,0)&=&R_y(-\omega_B t)R_{\rm tot}(\omega_{\rm tot}({\bf k})t)\\
  &=&R_{x'(t)}(\beta_{\bf k}) R_y(\omega_{\rm eff}({\bf k})t)
  R_{x}(-\beta_{\bf k}),
  \label{rotation_matrix}
\end{eqnarray}
in which 
\begin{eqnarray}
  && \hspace{-0.4cm}
  \beta_{\bf k}\approx\tan\beta_{\bf k}=\Omega_z({\bf k})/\omega_B,\\
  && \hspace{-0.4cm}
  \omega_{\rm eff}({\bf k})
  =\sqrt{\omega_B^2+\Omega_z^2({\bf k})} -\omega_B
  \approx \Omega_z^2({\bf k})/(2\omega_B),\\
  && \hspace{-0.4cm}
  {\bf e}_{x'(t)}=R_y(-\omega_B t){\bf e}_{x}=
  \cos(\omega_B t){\bf e}_x+\sin(\omega_B t){\bf e}_z.
\end{eqnarray}
In the above derivation, we have used the theorem 
$R_j(\theta_j)R_i(\theta_i)R_j(-\theta_j)=R_{i'}(\theta_i)$
with ${\bf e}_{i'}=R_j(\theta_j){\bf e}_{i}$
and the condition $\omega_B\gg \langle|\Omega_z({\bf k})|\rangle$.
We further limit ourselves in the regime 
$\langle|\omega_{\rm eff}({\bf k})|\rangle t \sim 
\langle|\omega_{\rm eff}({\bf k})|\rangle \tau_p \ll 1$,\cite{valid} 
thus all relevant
rotation angles in Eq.~(\ref{rotation_matrix}) are very small and the
corresponding rotation vectors satisfy the vector summation rule. 
Then one obtains the rotation vector $\bm{\theta}_{\bf k}(t,0)$, which
corresponds to $U_{\bf k}(t,0)=\exp[-i{\bf J}\cdot \bm{\theta}_{\bf k}(t,0)]$ 
with ${\bf J}$ representing the angular momentum operator and 
\begin{eqnarray}
  \nonumber
  \bm{\theta}_{\bf k}(t,0)&=& \omega_{\rm eff}({\bf k})t\, {\bf e}_y
  + \beta_{\bf k} [\cos(\omega_B t)-1] {\bf e}_x \\
  && {}+ \beta_{\bf k} \sin(\omega_B t) {\bf e}_z.
  \label{rotation_0}
\end{eqnarray}
The above equation can also be understood with the help of 
Fig.~\ref{fig_scheme}.
The first two terms in Eq.~(\ref{rotation_0}) just correspond to the 
angle between $\tilde{\bf S}_{{\bf k}}(0)$ 
[$\tilde{\bf S}_{{\bf k}}(0)={\bf S}_{{\bf k}}(0)$] and
the projection of $\tilde{\bf S}_{{\bf k}}(t)$ in the $x$-$z$ plane
($\theta_1$) and the angle between $\tilde{\bf S}_{{\bf k}}(t)$ and the
$x$-$z$ plane ($\theta_2$), respectively, illustrated  in Fig.~\ref{fig_scheme},
while the third term is ineffective since $\tilde{\bf S}_{{\bf k}}(0)\parallel
z$ in this case.

As mentioned above, all the relevant rotation angles are very small, 
thus the rotation vector between two adjacent scattering events occuring 
at $t$ and $t'$ reads
$\bm{\theta}_{\bf k}(t,t')=\bm{\theta}_{\bf k}(t,0)-\bm{\theta}_{\bf k}(t',0)$.
Averaging over $\tau=t-t'$ and $T=(t+t')/2$, one obtains the mean square of the
rotation angle between two adjacent scattering events in the case with the
initial spin vector along ${\bf e}_i$, 
\begin{equation}
  \theta_{{\rm ave},i}^2({\bf k}) = \int_{0}^{\tau_p} \frac{d\tau}{\tau_p}
  \int_0^{T_B} \frac{dT}{T_B} \Big|\overline{\bm{\theta}_{{\bf k}}(\tau,T)} 
  -\overline{\bm{\theta}_{{\bf k}}(\tau,T)}\cdot {\bf e}_i{\bf e}_i\Big|^2,
  \label{rotation_angle_ave}
\end{equation}
in which $T_B=2\pi/\omega_B$.
Considering $T_B\ll\tau_p$, Eq.~(\ref{rotation_angle_ave}) reads
\begin{eqnarray}
  \nonumber
  \theta_{{\rm ave},z}^2({\bf k}) &=& \frac{1}{\tau_p} \int_{0}^{\tau_p} d\tau\,
  \overline{\omega_{\rm eff}({\bf k})}^2 \tau^2 + \frac{4\beta_{\bf k}^2}{T_B^2} \\
  \nonumber
  && \hspace{-0.cm} {} \times \int_0^{T_B} d\tau \int_0^{T_B} dT\;
  \sin^2(\omega_B \frac{\tau}{2}) \sin^2(\omega_B T) \\
  &=& \overline{\omega_{\rm eff}({\bf k})}^2 \tau_p^2/3
  + \beta_{\bf k}^2,
  \label{rotation_z_ave} \\
  \theta_{{\rm ave},x}^2({\bf k})&=& \overline{\omega_{\rm eff}({\bf k})}^2
  \tau_p^2/3 + \beta_{\bf k}^2, 
  \label{rotation_x_ave} \\
  \theta_{{\rm ave},y}^2({\bf k})&=& 2\beta_{\bf k}^2.
  \label{rotation_y_ave}
\end{eqnarray}
Further exploiting the approximate formula of the DP SRT based
on the random walk theory,\cite{opt-or} 
\begin{equation}
  \tau_{s,i}^{-1}\sim \langle {\theta}_{{\rm ave},i}^2 ({\bf k})
  \rangle \tau_p^{-1},
  \label{SRT_estimate}
\end{equation} 
one obtains the SRTs given by Eqs.~(\ref{SRT_z_rough}) and (\ref{SRT_y_rough}). 
From the above discussions, one finds that the terms 
$\overline{\omega_{\rm eff}({\bf k})}^2 \tau_p^2$ and $\beta_{\bf k}^2$ in 
Eqs.~(\ref{rotation_z_ave})-(\ref{rotation_y_ave}) describe two kinds of 
inhomogeneous broadening, which induce the DP-like ($\tau_s\propto
\tau_p^{-1}$) and EY-like ($\tau_s\propto \tau_p$) behaviors,
respectively. This is exactly the cause of the anomalous $\tau_s$-$\tau_p$
relations of the DP mechanism under a strong in-plane magnetic field.

\section{Investigations via KSBEs}  
In order to obtain the exact SRT, we turn to the fully microscopic KSBE
approach.\cite{wu_review}
As mentioned in the introduction, we choose the investigated system to be the
symmetric (110) 
QWs.\cite{Ohno_110,Wu_110,Dohrmann_04,spin_noise_110,Sherman_110}
In fact, similar results can be obtained in (100) QWs with identical 
Dresselhaus and Rashba SOC strengths\cite{Averkiev_identical,Cheng} and are not
repeated here. The KSBEs can be written as 
\begin{equation}
  \partial_t{\rho}_{{\bf k}}= -i\left[\omega_B \frac{{\sigma}_y}{2} 
  + \bm{\Omega}({\bf k}) \cdot \frac{\bm{\sigma}}{2},\;\; {\rho}_{{\bf k}} \right] 
  +\left.\partial_t{\rho}_{{\bf k}}\right|_{\rm scat},
\end{equation}
in which $[\ ,\ ]$ denotes the commutator and
${\rho}_{\bf k}$ represents the density matrix of electron with momentum
${\bf k}$. The scattering term $\left.\partial_t{\rho}_{{\bf k}}\right|_{\rm scat}$
consists of the electron-impurity, electron--longitudinal-optical-phonon,
electron--acoustic-phonon and electron-electron Coulomb scatterings with their
expressions given in detail in Ref.~\onlinecite{Zhou_PRB_07}.

\subsection{Analytic study}
Before discussing the numerical results by solving the KSBEs,
we first investigate the spin relaxation analytically in a simplified case, where 
only the linear-$k$ term in the Dresselhaus SOC and the elastic
scattering (i.e., the electron-impurity scattering) are retained.
Transforming the density matrix into the interaction picture as
$\tilde{\rho}_{{\bf k}}=e^{iH_B t}\rho_{{\bf k}}e^{-iH_B t}$ and 
defining the spin vector 
\begin{align}
  \tilde{\bf S}_{k,l}=
  {\rm Tr}\big[\tilde{\rho}_{k,l}\bm{\sigma}\big], \qquad
  \tilde{\rho}_{k,l}=\frac{1}{2\pi}\int_0^{2\pi} d\phi_{\bf k} 
  \tilde{\rho}_{\bf k} e^{il\phi_{\bf k}},
\end{align}
one obtains 
\begin{align}
  & \partial_t\tilde{\bf S}_{{\bf k},l}(t)= \tilde{U}_{\rm so}(t) 
  \tilde{\bf S}_{{\bf k},l\pm 1}(t)
  -{\tilde{\bf S}_{{\bf k},l}(t)}/{\tau_{p,l}}.
  \label{S_angle_int}
\end{align}
Here
\begin{align}
  & \tilde{U}_{\rm so}(t) = \Omega_{\rm so} \begin{pmatrix} 
    0 & -\cos\omega_B t & 0 \\ 
    \cos\omega_B t & 0 & \sin\omega_B t \\
    0 & -\sin\omega_B t & 0 \end{pmatrix}
\end{align}
in which $\Omega_{\rm so}=-{\gamma_D}\langle k_z^2\rangle_0 k/4$ and
\begin{align}
  {\tau_{p,l}^{-1}}= \frac{N_i}{2\pi} \int_0^{2\pi} d\phi_{\bf k} \;
  |W_{\rm ei}(k,\phi_{\bf k})|^2 (1-\cos l\phi_{\bf k})
\end{align}
(note that $\tau_{p,1}=\tau_{p}$) with $W_{\rm ei}(k,\phi_{\bf k})$ standing for
the matrix element of the electron-impurity scattering. 
Retaining terms with $|l|\le 2$, Eq.~(\ref{S_angle_int}) can be reduced into
\begin{align}
  \nonumber
  \partial_t \tilde{\bf S}_{{\bf k},0}(t) &= e^{-t/\tau_p}\tilde{U}_{so}(t)
  \int_0^t dt_1 e^{t_1/\tau_p} \tilde{U}_{so}(t_1) 
  \Big[ 2 \tilde{\bf S}_{{\bf k},0}(t_1)  \\
  & {} + e^{-t_1/\tau_{p,2}} \int_0^{t_1} dt_2 e^{t_2/\tau_{p,2}}
  \partial_{t_2} \tilde{\bf S}_{{\bf k},0}(t_2) \Big].
  \label{exact_form}
\end{align}
Next, we replace $\tilde{\bf S}_{{\bf k},0}(t_1)$ by 
$\tilde{\bf S}_{{\bf k},0}(t)$ following the Markovian approximation and
transform the above equation into the iterate form,
\begin{align}
  \partial_t \tilde{\bf S}_{{\bf k},0}(t) = & \Big\{\Gamma_1(t) 
  + {\cal F}[\Gamma_1(t)] + {\cal F}[{\cal F}[\Gamma_1(t)]] + ... \Big\} 
  \tilde{\bf S}_{{\bf k},0}(t), 
  \label{iterate_form}
  \\
  \Gamma_1(t) = &\; 2 e^{-t/\tau_p}\tilde{U}_{so}(t) \int_0^t dt_1 e^{t_1/\tau_p}
  \tilde{U}_{so}(t_1), \\
  \nonumber
  {\cal F}[g(t)] = &\; e^{-t/\tau_p}\tilde{U}_{so}(t) \int_0^t dt_1 e^{t_1/\tau_p}
  \tilde{U}_{so}(t_1) \\
  & \times e^{-t_1/\tau_{p,2}} \int_0^{t_1} dt_2 e^{t_2/\tau_{p,2}} g(t_2).
\end{align}
Further considering the magnitude of $\tilde{U}_{so}(t)$
is much smaller than its frequency $\omega_B$, we apply the rotating-wave
approximation and only retain the terms with time-independent coefficients on
the right side of Eq.~(\ref{iterate_form}). 
Then one obtains the SRTs, to the leading order, 
\begin{align}
  & \tau_{sz}^{-1}=\tau_{sx}^{-1}= \left\langle
  \frac{\overline{\Omega_z^2({\bf k})}^2 \tau_{p,2}}{4\omega_B^2} 
  +\frac{\overline{\Omega_z({\bf k})}^2\tau_{p}}
  {2(1+\omega_B^2\tau_{p}^2)} \right\rangle,
  \label{SRT_z} \\
  & \tau_{sy}^{-1}= \left\langle \frac{\overline{\Omega_z({\bf k})}^2\tau_{p}}
    {1+\omega_B^2\tau_{p}^2} \right\rangle.
  \label{SRT_y} 
\end{align}
Note that the factor $1/(1+\omega_B^2\tau_{p}^2)$ also appears in the
previous works on the spin relaxation under a magnetic
field.\cite{Ivchenko_mag,Margulis_mag,Burkov_mag,Glazov_mag} 
In the strong-magnetic-field limit, $\omega_B\tau_{p}\gg 1$,
one recovers Eqs.~(\ref{SRT_z_rough}) and (\ref{SRT_y_rough}) from the above 
equations with $a=\tau_{p,2}/\tau_{p}$ and $b=b'=1/2$.
In addition, after considering the correction of the cubic Dresselhaus term, 
$\overline{\Omega_z({\bf k})}$ and $\overline{\Omega_z^2({\bf k})}$ in 
the above equations should be replaced by 
\begin{equation}
  \left.\Omega_z({\bf k})\right|_{l=1}=\frac{\gamma_D k}{8}
  \left({k^2}-4\langle k_z^2\rangle_0\right)\cos\phi_{\bf k}
\end{equation}
and
\begin{align}
  \left.\Omega_z^2({\bf k})\right|_{l=2}= \frac{\gamma_D^2 k^2}{128}
  \left({k^2}-4\langle k_z^2\rangle_0\right)
  \left(7{k^2}-4\langle k_z^2\rangle_0\right) \cos2\phi_{\bf k},
\end{align}
respectively.

\subsection{Numerical results}
In this subsection, we investigate the exact SRT by numerically solving the
KSBEs with all the scatterings explicitly included.
We choose a symmetric (110) InAs QW due to its large Land\'e $g$ factor. 
For this material, $g=-14.3$\cite{Rocca_g_factor} and
$\gamma_D=-27.8$~eV~\AA$^3$.\cite{Kotani_gamma_D} 
The other material parameters can be found in Ref.~\onlinecite{para_semi}.
We further set $B=2$ or $4$~T, both satisfying the condition
$\omega_{B}\gg \langle|{\Omega_z({\bf k})}|\rangle$.
The well width is chosen to be $a=5$~nm, which is smaller than the
cyclotron radius of the lowest Landau level so that the orbital  
effect from the external magnetic field is irrelevant.
It is noted that both the magnetic field and well width chosen here are within
the experimental feasibility.
In addition, the electron density is chosen to be $N_e=3\times
10^{11}$~cm$^{-2}$. The corresponding Fermi energy $E_{\rm F}\approx 30$~meV,
which is much larger than the Zeeman splitting about $3$~meV for $B=4$~T.
Therefore, the inclusion of the Zeeman splitting in 
the energy-conservation delta functions in the scattering terms of the KSBEs 
is unimportant to the relaxation of the out-of-equilibrium spin
polarization we investigate.\cite{Cheng,Grimaldi}

\begin{figure}[tbp]
  \begin{center}
    \includegraphics[width=7.cm]{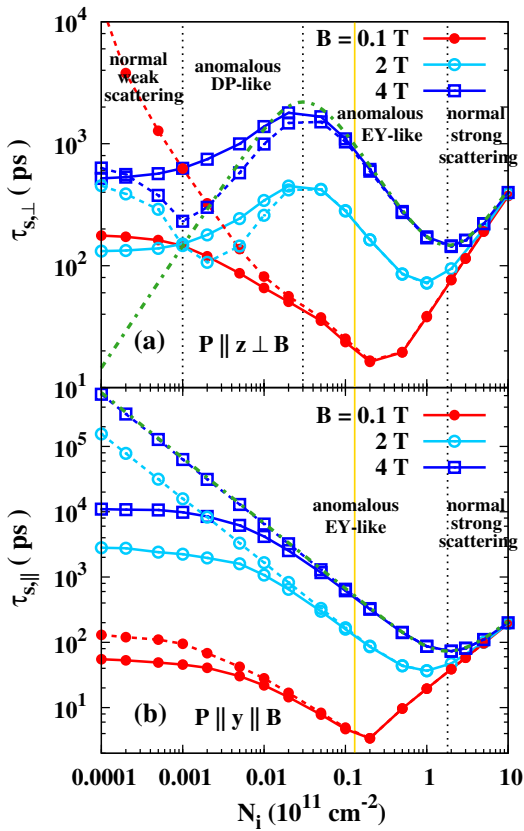}
  \end{center}
  \caption{(Color online) Transverse (a) and longitudinal (b) SRTs 
    from the calculations with all the relevant
    scatterings (solid curves) and only the electron-impurity scattering
    (dashed curves) against the impurity density under different magnetic fields.
    The green chain curves in (a) and (b) are the results from
    Eqs.~(\ref{SRT_z}) and (\ref{SRT_y}), respectively.
    The vertical black dotted lines indicate the boundaries between different
    regimes under the strong magnetic field $B=4$~T.
    The vertical yellow lines indicate the boundary between the weak
    and strong scattering regimes in the weak field case.
  }
  \label{fig_impurity} 
\end{figure}

We first compare the SRTs from the KSBEs with only the electron-impurity
scattering with those from Eqs.~(\ref{SRT_z}) and (\ref{SRT_y}) with
the correction of the cubic Dresselhaus term.
In Fig.~\ref{fig_impurity}, the results from these two approaches are plotted as 
the blue dashed and green chain curves for $B=4$~T in the case with the
initial ensemble average spin polarization 
${\bf P}=\sum\limits_{\bf k}{\bf S}_{\bf k}(0)/N_e$ 
along the $z$- and $y$-axes (recall ${\bf B}\parallel y$). 
The temperature and initial spin polarization are chosen to be
$T=15$~K and $|{\bf P}|=0.1$~\%. 
It is shown that the results from the numerical computations agree fairly well
with the approximate formula in the impurity density regime satisfying 
$\tau_{p,2}^{-1}>\langle|\overline{\omega_{\rm eff}({\bf k})}|\rangle$ for the
transverse SRT [${\bf P}\perp{\bf B}$, Fig.~\ref{fig_impurity}(a)], and in the
whole impurity density regime for the 
longitudinal SRT [${\bf P}\parallel {\bf B}$,
Fig.~\ref{fig_impurity}(b)]. This further justifies the validity of 
Eqs.~(\ref{SRT_z}) and (\ref{SRT_y}) in these regimes, in
consistence with the discussions presented above.\cite{tau_2}

From this figure, one also observes that the behaviors of the DP spin relaxation
under a strong magnetic field (blue dashed curves) are very different from the
conventional ones under a weak magnetic field (red solid curves).
We first focus on the transverse SRT [Fig.~\ref{fig_impurity}(a)]. 
It is seen that the spin relaxation in this case can be divided into four
regimes (separated by the vertical black dotted lines),
in contrast to the two regimes in the weak field case, i.e., 
the weak and strong scattering regimes (separated by the
vertical yellow solid line). The two regimes in the middle 
[$\tau_{p,2}\langle|\overline{\omega_{\rm eff}
({\bf k})}|\rangle/\tau_{p}<\tau_p^{-1}<\omega_B$] 
are just the anomalous DP- and EY-like regimes
discussed above, respectively.
It is shown that, in the anomalous DP-like regime, which is in the
original weak scattering limit, the DP SRT shows the
strong scattering behavior ($\tau_s\propto\tau_p^{-1}$).  
Moreover, in the anomalous EY-like regime, most of which is in the original
strong scattering limit, the DP SRT exhibits the EY-like behavior
($\tau_s\propto\tau_p$). 
All these anomalous behaviors come from the unique form of the inhomogeneous
broadening given by Eq.~(\ref{rotation_z_ave}), just as discussed above.
A peak appears at the boundary between these two regimes, i.e., $\tau_p^{-1}=
\sqrt{{\tau_{p,2}}\langle\overline{\Omega_z^2({\bf k})}^2\rangle/(2\tau_{p}
\langle{\Omega_z^2({\bf k})}\rangle)}$, which is independent
of the magnetic field. This effect comes from the competition of the
two kinds of inhomogeneous broadening in Eq.~(\ref{rotation_z_ave}).
For lower (higher) impurity densities beyond the above regimes, 
the SRT exhibits the conventional DP behavior in the weak (strong)
scattering limit. Thus we refer to these two regimes as the normal weak and
strong scattering regimes, respectively. The behavior in the normal weak
scattering regime can be understood by considering that 
the impurity scattering is too weak to suppress the inhomogeneous broadening
from $\omega_{\rm eff}({\bf k})$ in Eq.~(\ref{rotation_z_ave}), similar to the
conventional weak scattering case.
As for the normal strong scattering regime, the underlying physics is that when
$\omega_B\tau_p\ll 1$, the inhomogeneous broadening returns to the conventional
form, which can be demonstrated by exploiting
Eq.~(\ref{rotation_angle_ave}) and considering 
$\sin(\omega \tau/2)\approx \omega\tau/2$ for $0<\tau<\tau_p$.
We then turn to the longitudinal SRT [Fig.~\ref{fig_impurity}(b)]. 
There are only two regimes in this case. In the regime $\tau_p^{-1}<\omega_B$, the SRT
decreases with increasing $N_i$, which comes from the inhomogeneous broadening
given by Eq.~(\ref{rotation_y_ave}), similar to the anomalous EY-like regime for
the transverse SRT.
In the regime $\tau_p^{-1}>\omega_B$, the SRT increases with $N_i$, because the
inhomogeneous broadening returns to the conventional form, just as the normal
strong scattering regime for the transverse SRT. 
Note that there is no normal weak scattering regime in this case. This is 
because when ${\bf P}\parallel y$, the term $R_y(\omega_{\rm eff}({\bf k})t)$ in
the rotation matrix [Eq.~(\ref{rotation_matrix})] does not contribute to
the rotation angle\cite{rot_y} and the corresponding rotation angles between
adjacent scattering events become independent of $\tau_p$.

In Fig.~\ref{fig_impurity}, we also plot the SRT with only the impurity
scattering for $B=2$~T as the azure dashed curve. 
It is shown that the SRT in this case is shorter than the corresponding one for
$B=4$~T in the anomalous DP- and EY-like regimes but becomes very close to
the latter one in the normal strong scattering regime, all of which are
consistent with the form of the inhomogeneous broadening discussed above. 
It is also seen that in the case of ${\bf P}\perp{\bf B}$
[Fig.~\ref{fig_impurity}(a)], the areas of both the anomalous DP- and EY-like
regimes for $B=2$~T are smaller than those for $B=4$~T, while the positions of
the peak remain fixed.
These behaviors are consistent with the above discussions on the boundaries
between different regimes. 

Then we discuss the SRTs with all the relevant scatterings. The results 
are plotted as the solid curves in Fig.~\ref{fig_impurity}.
One observes that the behaviors in these cases are similar to the corresponding
ones with only the electron-impurity scattering, especially all the anomalous
behaviors in the anomalous DP- and EY-like regimes are retained with all
scatterings included. This further justifies that these anomalous 
behaviors can be observed in experiments. 
It is also shown that the SRT with all scatterings is longer than that 
with only the impurity scattering in the anomalous DP-like regime for 
${\bf P}\perp{\bf B}$, while shorter than the latter one in the anomalous
EY-like regime for ${\bf P}\parallel{\bf B}$. All these behaviors are 
consistent with Eqs.~(\ref{SRT_z}) and (\ref{SRT_y}). 
In addition, it is seen that the normal weak scattering regime, which previously
appears at extremely low impurity for ${\bf P}\perp{\bf B}$, disappears in
the impurity density dependence with all scatterings. 
This is because the condition
$\tau_{p,2}^{-1}>\langle|\overline{\omega_{\rm eff}({\bf k})}|\rangle$ is always
satisfied due to the inclusion of the other scatterings.

\begin{figure}[tbp]
  \begin{center}
    \includegraphics[width=7.cm]{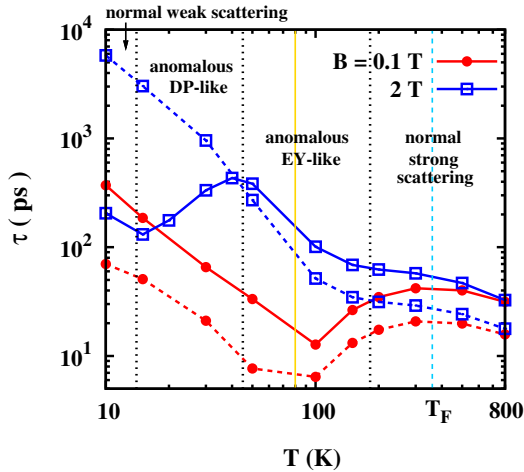}
  \end{center}
  \caption{(Color online) Transverse (solid curves) and longitudinal (dashed
    curves) SRTs against the temperature for $N_i=0$ under different magnetic
    fields. The vertical azure dashed line indicates the temperature satisfying
    $T=T_{\rm F}$. The vertical black dotted lines separate the boundaries
    between different regimes for transverse SRT under the strong magnetic field
  $B=4$~T. The vertical yellow line illustrates the boundary between the weak
and strong scattering regimes in the case of  weak magnetic field.}
  \label{fig_T} 
\end{figure}

The anomalous scaling of the DP spin relaxation also significantly
influences the temperature dependence of the SRT.
The transverse and longitudinal SRTs are plotted as function of temperature for
$N_i=0$ under different magnetic fields in Fig.~\ref{fig_T}. 
We first focus on the transverse SRT.
The SRT under strong magnetic field first exhibits a valley and then a peak.
The underlying physics is as follows. 
In the degenerate limit (i.e., 
$T\ll T_{\rm F}=E_{\rm F}/k_{\rm B}$ with $T_{\rm F}\approx 360$~K here),
both the electron-electron and electron-phonon scatterings increase with 
increasing temperature, while the inhomogeneous broadening is insensitive to the
temperature. Thus, the temperature dependence of the SRT is just determined by
the momentum relaxation.
As shown in Fig.~\ref{fig_T}, all the normal weak scattering, anomalous DP-like
and EY-like regimes are in the degenerate limit. Therefore, the peak and
valley appear around the boundaries between these three regimes, just similar to
the impurity density dependence discussed above.
However, it is also seen that no valley appears in the temperature dependence
around the boundary between the anomalous EY-like and normal strong scattering
regimes. This is because most of the normal strong scattering regime is in the
nondegenerate limit ($T\gg T_{\rm F}$).
In this limit, the scattering becomes insensitive to the temperature due to the
competition of the decrease of the electron-electron scattering and the increase
of the electron-phonon scattering, whereas the inhomogeneous broadening
increases rapidly with the temperature. Thus, the SRT decreases with temperature
in this regime. This leads to the absence of the valley.
The SRT under weak magnetic field also shows first a valley and then a peak.
But the positions of these valley and peak are quite different from the previous
ones and the underlying physics is totally different. 
The valley can be understood by considering that the boundary between the weak and
strong scattering regimes is in the degenerate limit. The peak
is due to the crossover of the degenerate and nondegenerate limits in the strong
scattering regime, which is well known in the
literature.\cite{wu_review,Zhou_PRB_07}  
Then we turn to the longitudinal SRT. 
It is seen that the behaviors under weak magnetic field are very similar to the
corresponding transverse ones, but the behaviors under strong magnetic field
become quite different: the SRT decreases monotonically with temperature. This
is just because the system in this case belongs to the anomalous EY-like regime
at low temperature.

\section{Conclusion and Discussion}
In conclusion, we have investigated the anomalous scaling of the DP SRT with the
momentum relaxation time in semiconductor QWs under a strong magnetic field,
whose direction is parallel to the QW plane and perpendicular to the spin-orbit
field. 
We discover that, for the transverse SRT perpendicular to the magnetic field,
the anomalous scaling occurs at two regimes, i.e., the anomalous DP- and
EY-like regimes. In the anomalous DP-like regime, which is in the
original weak scattering limit, the DP SRT is inversely proportional to the
momentum relaxation time, i.e., the strong scattering behavior. 
On the other hands, in the anomalous EY-like regime, which is in the
original weak scattering limit, the DP SRT is proportional to the momentum
relaxation time, i.e, the EY-like behavior,
both in the {\em opposite} trends against the conventional DP ones. 
As for the longitudinal SRT parallel to the magnetic field, the DP SRT is always
proportional to the momentum relaxation time even in the original strong
scattering limit, similar to the anomalous EY-like regime for the transverse SRT.
We further demonstrate that all these anomalous scaling relations come from the
unique form of the effective inhomogeneous broadening. 

Finally, we address the choice of the material. In the above calculations, we
choose InAs (110) QWs. In fact, similar behaviors also appear in (110) QWs
made of the other materials with large Land\'e $g$ factor, e.g.,
InSb\cite{para_semi}, and (100) QWs with identical Dresselhaus and Rashba
strengths made of InAs and InSb. 
However, the situation becomes very different for QWs made of materials
with small $g$ factor, e.g., GaAs, since the conditions $\omega_{B}\gg
\langle|{\Omega_z({\bf k})}|\rangle$ and the well width is smaller than the 
cyclotron radius of the lowest Landau level cannot be satisfied simultaneously.

\begin{acknowledgments}
  This work was supported by the National Basic Research Program of China under
  Grant No.\ 2012CB922002 and the Strategic Priority Research Program of the
  Chinese Academy of Sciences under Grant No. XDB01000000. 
\end{acknowledgments}


\begin{thebibliography}{0}
\bibitem{tau_2} It is reasonable to replace $\tau_p$ in the original validity
  condition of Eq.~(\ref{SRT_z_rough}) by $\tau_{p,2}$, as $\tau_{p,2}$ is
  exactly the momentum relaxation time related to the inhomogeneous broadening
  from $\omega_{\rm eff}$, shown in the angular expansion of the KSBEs.
\end{thebibliography}
\end{document}